\documentclass{article}

\usepackage{arxiv}

\usepackage{amsmath}
\usepackage[utf8]{inputenc} % allow utf-8 input
\usepackage[T1]{fontenc}    % use 8-bit T1 fonts
\usepackage{hyperref}       % hyperlinks
\usepackage{url}            % simple URL typesetting
\usepackage{booktabs}       % professional-quality tables
\usepackage{amsfonts}       % blackboard math symbols
\usepackage{nicefrac}       % compact symbols for 1/2, etc.
\usepackage{microtype}      % microtypography
\usepackage{cleveref}       % smart cross-referencing
\usepackage{lipsum}         % Can be removed after putting your text content
\usepackage{graphicx}
\usepackage{natbib}
\usepackage{doi}
\usepackage{algorithm}      % 提供最外层的 float 环境
\usepackage{algpseudocode}  % 提供 \Require, \Ensure, \Statex 等命令
\usepackage{siunitx}   % 解决 'S' 列报错 (Illegal character)
\usepackage{multirow}  % 解决 \multirow 报错
\usepackage{enumitem}

\graphicspath{{figs/}} % 设置图片路径

\title{LEAPS: An LLM-Empowered Adaptive Plugin in Taobao AI Search}

\newif\ifuniqueAffiliation
\uniqueAffiliationtrue

\ifuniqueAffiliation % Standard variant of author block
\author{{Lei Wang}\thanks{Equal contribution} \\
	Taobao \& Tmall Group of Alibaba\\
	Hangzhou 311121, China \\
	\texttt{yufang.yf123@alibaba-inc.com} \\
	\And
	{Jinhang Wu}\footnotemark[1] \\
	Taobao \& Tmall Group of Alibaba\\
	Hangzhou 311121, China \\
	\texttt{jinhang.wjh@taobao.com} \\
	\And
	{Zhibin Wang} \\
	Taobao \& Tmall Group of Alibaba\\
	Hangzhou 311121, China \\
	\texttt{wzb395642@taobao.com} \\
	\And
	{Biye Li}\thanks{Corresponding author} \\
	Taobao \& Tmall Group of Alibaba\\
	Hangzhou 311121, China \\
	\texttt{libiye.lby@alibaba-inc.com} \\
}
\else
\usepackage{authblk}

\newbox{\orcid}\sbox{\orcid}{\includegraphics[scale=0.06]{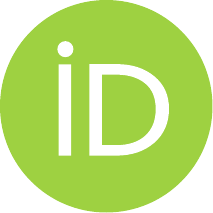}} 
\author[1]{%
	\href{https://orcid.org/0000-0000-0000-0000}{\usebox{\orcid}\hspace{1mm}David S.~Hippocampus\thanks{\texttt{hippo@cs.cranberry-lemon.edu}}}%
}
\author[1,2]{%
	\href{https://orcid.org/0000-0000-0000-0000}{\usebox{\orcid}\hspace{1mm}Elias D.~Striatum\thanks{\texttt{stariate@ee.mount-sheikh.edu}}}%
}
\affil[1]{Department of Computer Science, Cranberry-Lemon University, Pittsburgh, PA 15213}
\affil[2]{Department of Electrical Engineering, Mount-Sheikh University, Santa Narimana, Levand}
\fi

\renewcommand{\shorttitle}{LEAPS: An LLM-Empowered Adaptive Plugin in Taobao AI Search}

\hypersetup{
pdftitle={LEAPS: An LLM-Empowered Adaptive Plugin in Taobao AI Search},
pdfsubject={cs.IR, cs.CL, cs.AI},
pdfauthor={Lei Wang, Jinhang Wu, Zhibin Wang, Biye Li},
pdfkeywords={Conversational Product Search, Query Rewriting, Large Language Models, Reinforcement Learning},
}

\begin{document}
\maketitle

\begin{abstract}
	The rapid rise of large language models has shifted user search behavior from discrete keywords to natural-language, multi-constraint queries---a shift existing e-commerce search architectures struggle to accommodate. Users face a dilemma: precise natural-language queries often trigger zero-result scenarios, while forced simplification yields noisy, generic results that overwhelm decision-making. To address this, we propose \textbf{LEAPS} (\textbf{L}LM-\textbf{E}mpowered \textbf{A}daptive \textbf{P}lugin in Taobao AI \textbf{S}earch), which upgrades traditional search pipelines via a ``Broaden-and-Refine'' paradigm by attaching plugins at both ends. (1) Upstream, a \textit{Query Expander} generates adaptive, complementary query combinations to maximize the candidate set, trained via a three-stage strategy of inverse data augmentation, posterior-knowledge supervised fine-tuning, and diversity-aware reinforcement learning. (2) Downstream, a \textit{Relevance Verifier} filters noise by synthesizing multi-source signals (e.g., OCR text, reviews) with chain-of-thought reasoning. Extensive offline experiments and online A/B testing show that LEAPS significantly enhances the conversational shopping experience, while its non-intrusive architecture preserves established short-text retrieval performance and enables low-cost integration with diverse back-ends. Fully deployed on Taobao AI Search since August 2025, LEAPS serves hundreds of millions of users monthly.
\end{abstract}

% keywords can be removed
% \keywords{First keyword \and Second keyword \and More}

\section{Introduction}
\label{sec:introduction}

As Large Language Models (LLMs) reshape the paradigm of digital interaction~\cite{DBLP:journals/corr/abs-2501-12948,DBLP:conf/nips/Ouyang0JAWMZASR22}, e-commerce search stands at a critical crossroads. In this technological context, user search cognition is rapidly evolving; users are no longer satisfied with assembling discrete keywords but instead incline towards natural-language queries to articulate complex shopping needs~\cite{DBLP:journals/sigir/KuziM24,DBLP:conf/ijcai/IsyanbaevM25}.\footnote{Throughout this paper, we use the term \textit{conversational-style} to denote single-turn, natural-language queries that integrate multiple attribute constraints, scenario descriptions, and negation logic, as opposed to the multi-turn dialogue setting studied in conversational recommendation systems.} For instance, a user might input: \textit{``I want a linen blazer for men, 500-800 yuan, suitable for a beach wedding, no pure black, and no large patterns''}---a complex instruction integrating explicit attributes, numerical ranges, colloquial scenarios, and negation logic. Users expect AI-era search engines to act as intelligent shopping guides, interpreting such descriptions to deliver precisely matched items.

However, existing industrial search architectures reveal intrinsic capability limitations when addressing such high-dimensional conversational demands. Although mainstream hybrid retrieval architectures have integrated the strengths of sparse~\cite{DBLP:conf/sigir/FormalPC21} and dense retrieval~\cite{DBLP:conf/kdd/HuangSSXZPPOY20}, they remain inadequate when processing highly unstructured semantics with strict multi-attribute conjunctive conditions---frequently resulting in zero-result scenarios. Anticipating this limitation, users are compelled to artificially simplify their fine-grained needs into broad keyword combinations (e.g., \textit{``linen blazer men''}). While this compromise avoids null retrieval, it transfers the burden of filtering from the system to the user, who must navigate massive results filled with noise---items exceeding budgets, incorrect colors, or mismatched styles---leading to severe decision overload. Resolving the dichotomy between zero-result outcomes from precise descriptions and decision overload from generalized search has thus become a core proposition for AI-era e-commerce search.

Facing this dilemma, completely overhauling the underlying search architecture is not only cost-prohibitive but also fraught with risk---radical reconstruction threatens to disrupt existing retrieval experiences that are highly optimized for high-frequency short queries. Consequently, there is an urgent need to build a non-intrusive, adaptive plugin to bridge complex conversational-style intent with robust traditional search engines. To this end, this paper proposes the LEAPS (\textbf{L}LM-\textbf{E}mpowered \textbf{A}daptive \textbf{P}lugin in Taobao AI \textbf{S}earch) framework. Unlike a disruptive end-to-end overhaul, LEAPS adopts a modular Broaden-and-Refine paradigm, attaching intelligent plugins to both ends of the existing search pipeline. Upstream, a Query Expander converts complex user language into adaptive and complementary query combinations to maximize item coverage and resolve the zero-result issue. Downstream, a Relevance Verifier leverages LLM reasoning to prune noise that fails to meet complex constraints, resolving decision overload.

Although LLMs have advanced query rewriting~\cite{DBLP:conf/acl/NguyenMYAC25} and relevance filtering~\cite{DBLP:conf/sigir/WeiJCXL24,DBLP:conf/kdd/0001TS0K24,DBLP:conf/kdd/WangXZZNTWM24}, effectively integrating them into the complex ecosystem of e-commerce search remains a non-trivial challenge. Existing query rewriting methods often prioritize semantic similarity over search effectiveness~\cite{DBLP:journals/corr/abs-2305-14283}, while RL approaches relying on user interaction logs suffer in cold-start and long-tail scenarios. Relevance filtering models typically rely on surface-level text~\cite{DBLP:conf/sigir/LiuCXL0X24,DBLP:conf/sigir/MacAvaneyS23}, overlooking critical heterogeneous information like image text and reviews; combined with constrained reasoning capabilities~\cite{DBLP:conf/nips/Wei0SBIXCLZ22,DBLP:journals/corr/abs-2308-07107}, they struggle to effectively simulate complex human decision-making.

To address these challenges, LEAPS introduces targeted innovations for AI product search. For the Query Expander, we devise a three-stage training strategy: (1) Inverse Data Augmentation to bridge semantic gaps; (2) Posterior-knowledge-based Supervised Fine-Tuning (SFT) for preliminary adaptation to the search system; and (3) RL to enhance adaptability and item relevance while penalizing redundancy to improve diversity. For the Relevance Verifier, we approximate human evaluation by integrating multi-source information---including OCR text, reviews, and sales metrics---and employing multi-instruction SFT with Chain-of-Thought (CoT) to bolster reasoning capabilities for accurate screening.

The main contributions of this paper are as follows:
\begin{itemize}
  \item We propose LEAPS, a framework tailored for conversational-style queries. It has been fully deployed in Taobao AI Search, serving hundreds of millions of users monthly.
  \item LEAPS features a non-intrusive and adaptive architecture, which ensures high portability across diverse search backends and offers significant cost-efficiency.
  \item We introduce a diversity-aware reward, providing a single dense and interpretable signal that simultaneously enforces per-rewrite precision and rewrite-set complementarity.
\end{itemize}

\section{Related Work}

\paragraph{Query Rewriting.}
Existing query rewriting paradigms fall into two streams. \textit{Discriminative} approaches~\cite{DBLP:conf/sigir/FormalPC21,DBLP:conf/sigir/KhattabZ20,DBLP:conf/emnlp/ZhengHHH0Y20,DBLP:conf/aaai/LiuZ0WJD020} formulate rewriting as a term selection task, but their dependence on external knowledge bases introduces semantic drift, and the Matthew Effect in search logs biases models toward high-frequency patterns, leaving long-tail queries underserved. \textit{Generative} approaches~\cite{DBLP:conf/acl/ChengMD24,DBLP:conf/sigir/KostricB24} leverage sequence-to-sequence architectures, yet two issues persist: optimization typically targets system-agnostic semantic similarity, and although recent RL methods incorporate interaction signals such as CTR, sparse feedback degrades performance in cold-start and long-tail regimes; meanwhile, conventional decoding strategies tend to produce semantically homogeneous candidates whose result sets heavily overlap. The most directly comparable industrial system is MiniELM~\cite{DBLP:conf/acl/NguyenMYAC25}, which combines distillation, SFT, and RL for adaptive rewriting; we reproduce it as our primary baseline in Section~\ref{sec:setup}.

\paragraph{Search Relevance.}
Modern relevance modeling has evolved from pre-trained encoders such as BERT~\cite{DBLP:conf/naacl/DevlinCLT19} and DeBERTa~\cite{DBLP:conf/iclr/HeLGC21}---which lack the reasoning depth required by complex e-commerce queries---to LLM-based judges that exploit emergent cognitive capabilities~\cite{DBLP:conf/naacl/QinJHZWYSLLMWB24,DBLP:conf/emnlp/0001YMWRCYR23}. However, two industrial bottlenecks remain. First, most LLM-based judges consume only surface-level metadata such as titles and attributes~\cite{DBLP:conf/sigir/0003WZC0RRR24,DBLP:conf/emnlp/SaadanyBAKOW24}, ignoring rich heterogeneous signals including OCR text and user reviews. Second, low-latency requirements force them into a discriminative mode that emits scalar scores~\cite{DBLP:conf/sigir/MaWYWL24,DBLP:conf/sigir/ZhuangZKZ24}, bypassing CoT reasoning and sacrificing explainability. LEAPS addresses both: it treats posterior search yield as a dense, system-aware training signal and explicitly models set-level utility for rewriting, while integrating multi-source signals (OCR, reviews, sales) and internalizing concise CoT for relevance judgment, all within strict latency budgets. Orthogonally to prior work, LEAPS is non-intrusive---interacting with the underlying engine only through query submission and result aggregation---making it portable across heterogeneous back-ends.

\section{Method}
\subsection{Framework Overview}
\label{sec:method_overview}

\begin{figure*}[htbp]
  \centering
  \includegraphics[width=\textwidth]{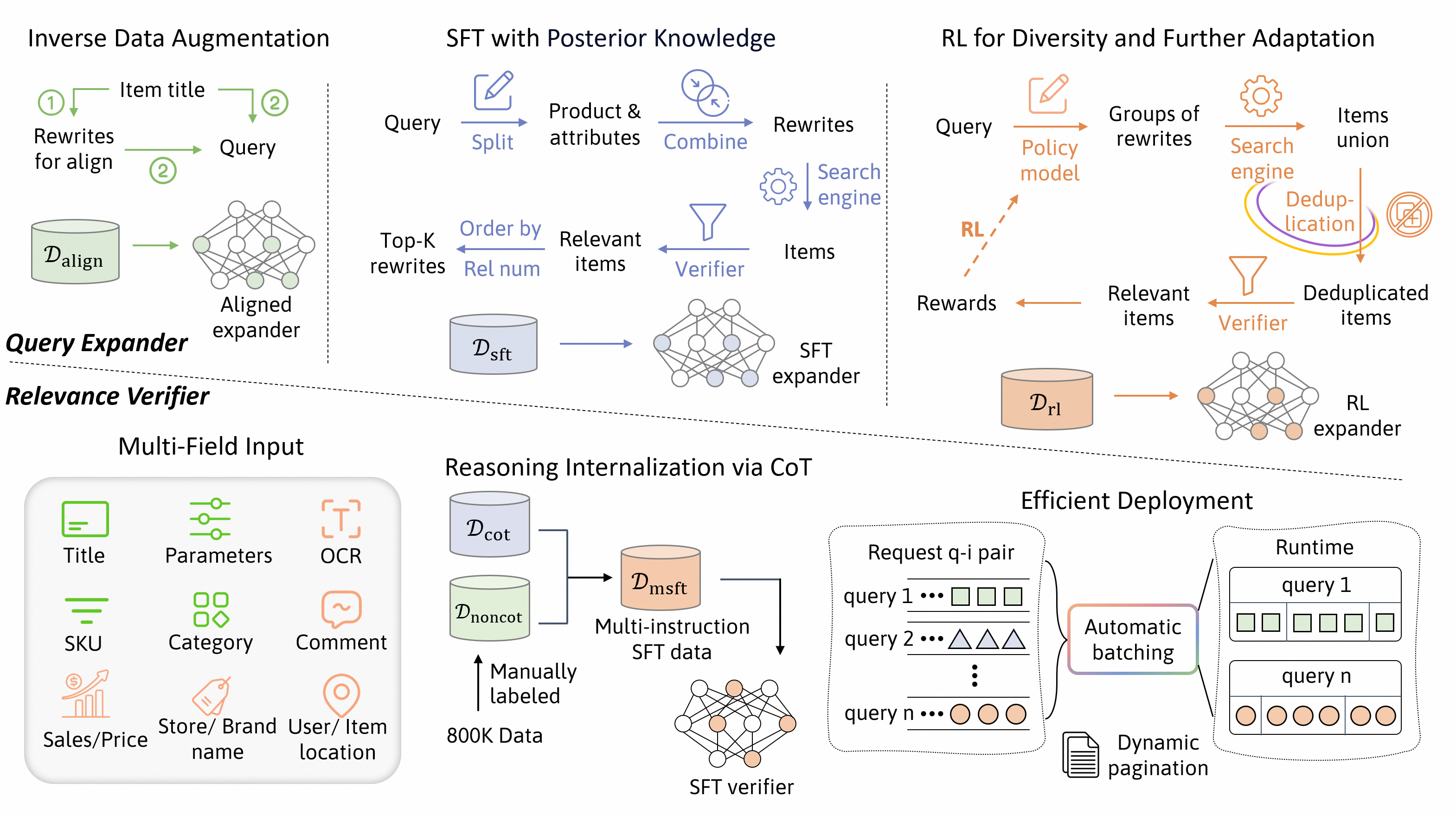}
  \caption{Schematic overview of the LEAPS framework. The Query Expander employs a three-stage training strategy; the Relevance Verifier synthesizes multi-source data with internalized CoT reasoning. LEAPS interacts with the search engine only via query submission and result aggregation.}
  \label{fig:framework}
\end{figure*}

The framework comprises two synergistic components: the \textit{Query Expander} and the \textit{Relevance Verifier}. Figure~\ref{fig:framework} provides a schematic illustration of the technical approach employed within these components.

The Query Expander ($\mathcal{E}$) is designed to increase the total number of potentially related items returned. $\mathcal{E}$ generates a set of rewrites $Q' = \{q^{\prime}_1, ..., q^{\prime}_n\}$ that are both complementary and system-friendly. In the execution phase, LEAPS submits $Q^{\prime}$ to the search engine $\mathcal{S}$, aggregating and deduplicating the returned results into a candidate pool $I_b$:
\begin{equation}
\label{eq:broaden}
  I_b = \bigcup_{q' \in \mathcal{E}(q)} \mathcal{S}(q').
\end{equation}

The Relevance Verifier ($\mathcal{V}$) functions as a discriminating agent, curating items that align with user intent from the potentially noisy set $I_b$. During inference, the Verifier incorporates personalized user profile $u$ to generate a binary relevance label accompanied by a concise rationale for each item $i \in I_b$ against the original query $q$. We denote the specific label indicating a positive match as $y_{\text{rel}}$, yielding the filtered item set $I_r$:
\begin{equation}
  \label{eq:verify}
    I_r = \{ i \in I_b \mid \mathcal{V}(q, i, u) = y_{\text{rel}} \}.
\end{equation}

Functionally, LEAPS treats the underlying search engine as a black box, interacting solely through standard query interfaces and result ingestion---a decoupled design enabling plug-and-play integration without access to internal indices or parameters. The dual plugins enhance search capabilities while keeping latency within conversational budgets: the Expander ($\sim$180 input / 20 output tokens) and the Verifier ($\sim$1{,}500 input / 12 output tokens) both stay within $\sim$290~ms at P99.

The two components are trained sequentially: we first train the Relevance Verifier $V$, then freeze it to train the Query Expander. Freezing $V$ eliminates circular dependencies, and its consistency with human annotation (validated in Section~\ref{sec:verifier}) bounds any bias it introduces as a training signal.

\subsection{Query Expander}

\subsubsection{Inverse Data Augmentation}

To bridge the lexical gap between colloquial queries and structured item indices, we adopt an Inverse Data Augmentation strategy that circumvents the scarcity of clean query-rewrite supervision in production logs: starting from curated product titles, we derive standardized rewrites and prompt DeepSeek-R1 to synthesize matching colloquial queries that mirror the register, informality, and constraint patterns of real Taobao AI Search logs. The resulting pairs are used for SFT under the standard negative log-likelihood objective:
\begin{equation}
  \mathcal{L}_{\text{align}}(\theta) = -\mathbb{E}_{(x,y) \sim \mathcal{D}_{\text{align}}} \sum_{i=1}^{|y|} \log \pi (y_i \mid y_{0:i-1}, x;\theta).
\end{equation}
This stage serves as a warm-up that instills domain-specific linguistic patterns.

\subsubsection{SFT with Posterior Knowledge}

Although alignment adapts the model to e-commerce language, complex multi-constraint queries often yield zero results when rigid matching fails on even one attribute. We hypothesize that query attributes have hierarchical importance: some are core determinants while others are auxiliary. For instance, in \textit{``quiet portable air conditioner''}, \textit{``portable air conditioner''} is the immutable core intent, whereas \textit{``quiet''} is a soft constraint inconsistently labeled in product metadata; indiscriminately enforcing it risks over-constraint. The goal is therefore to dynamically relax secondary constraints.

To this end, we adopt a ``generate-and-verify'' approach: we permute query attributes to form candidate rewrites, submit them to the search engine, and use the Verifier to measure the proportion of relevant products in the results. This search yield serves as posterior knowledge---a ground-truth proxy for rewrite effectiveness. The top-$k$ rewrites are selected for SFT (Algorithm~\ref{alg:sft_posterior}), with $k=4$ chosen from $\{3, 4, 5, 6\}$ on a 500-query development set for the best effectiveness--efficiency trade-off. At line~9, $u$ is a mocked profile (replaced by the real user's profile at serving time); a 500-query audit yields Spearman $\rho \approx 0.96$ between mocked- and real-profile rankings, confirming user-side features are largely orthogonal to rewrite-quality assessment. For the rare queries with $|A|>6$ ($\approx 3.2\%$ of training data), we merge semantically equivalent attributes and uniformly sample 50 subsets from $\mathcal{P}(A)$ as fallback, retaining 94\% of the exhaustive-enumeration F1@20 on a 500-query audit.

\begin{algorithm}
  \caption{Posterior-Knowledge-Based SFT Data Construction}
  \label{alg:sft_posterior}
  \begin{algorithmic}[1]
    \Require Original user query $q$, Aligned Expander $\pi_a$, Search Engine $\mathcal{S}$, Relevance Verifier $\mathcal{V}$, Truncation parameter $k$
    \Ensure Optimized SFT model $\pi_{\text{sft}}$
    \State $(p, A) \gets \pi_a(\text{prompt}_{\text{split}}, q)$ \Comment{Decompose $q$ into product $p$ and attribute set $A$}
    \State $Q_{\text{cand}} \gets \emptyset$
    \For{$A' \in \mathcal{P}(A)$} \Comment{$\mathcal{P}(\cdot)$: power set}
        \State $Q_{\text{cand}} \gets Q_{\text{cand}} \cup \{\textsc{Concat}(p, A')\}$
    \EndFor
    \State $\mathcal{U} \gets \emptyset$
    \For{$q' \in Q_{\text{cand}}$}
        \State $I \gets \mathcal{S}(q')$
        \State $I^{+} \gets \{i \in I : \mathcal{V}(q, i, u) = y_{\text{rel}}\}$ \Comment{$u$: mocked profile during training; real profile at serving}
        \State $s \gets |I^{+}| / |I|$ if $|I|>0$ else $0$
        \State $\mathcal{U} \gets \mathcal{U} \cup \{(s, q')\}$
    \EndFor
    \State $\{q'_1, \dots, q'_k\} \gets \textsc{TopK}(\textsc{Sort}(\mathcal{U}), k)$
    \State $\pi_{\text{sft}} \gets \operatorname{SFT}(\pi_a, \{(q, \text{Concat}(q'_1, \dots, q'_k))\})$
  \end{algorithmic}
\end{algorithm}

\subsubsection{RL for Query Diversification}
\label{sec:rl_for_query_adaptation}

SFT optimizes individual rewrites, but the Expander must produce a \emph{set} with collective coverage; posterior-SFT models collapse into high-probability modes, yielding redundant top-$k$ rewrites. We address this with a set-level \textit{Effective Relevance} reward.

Let $N$ be the per-query retrieval budget ($N{=}100$ in deployment) and $I_r$ the deduplicated set of verified-relevant items recovered within it:
\begin{equation}
  R_{\text{eff}} = |I_r| / N.
\end{equation}
Since $N$ is fixed and $I_r$ deduplicated, a rewrite duplicating its peers' results consumes budget without enlarging $I_r$, driving the Expander toward coverage rather than paraphrase. A stratified 5{,}000-pair audit (deduplicated against the Verifier's training corpus) shows Verifier judgments agreeing with trained annotators at ${\approx}95\%$ (Cohen's $\kappa{=}0.88$; bucket-wise $0.91/0.89/0.86/0.82$ on Simple/Moderate/Complex/Heavy queries)---the mild Heavy-bucket drop reflects intrinsic difficulty on multi-constraint inputs but stays in the substantial-agreement regime. To bound reward bias, we re-train the RL stage on a $1{,}000$-query subset with \emph{human-labeled} relevance as the reward; the resulting Expander differs from its Verifier-rewarded counterpart by only $0.01$ F1@20 ($0.82$ vs.\ $0.81$), confirming $R_{\text{eff}}$ as a faithful signal at a fraction of the labeling cost.

Queries with pre-RL expected $R_{\text{eff}}<0.1$ yield high-variance group rewards and those with $R_{\text{eff}}>0.9$ produce near-zero gradient; restricting the RL pool to $[0.1, 0.9]$ substantially stabilizes training.

We optimize with GRPO~\cite{DBLP:journals/corr/abs-2402-03300}: $R_{\text{eff}}$ is sparse and sequence-level, limiting the value head's benefit, while groupwise normalization over per-query rewrite sets aligns with our set-level objective. We use group size $8$ and KL coefficient $\beta{=}0.04$, selected on a 500-query dev set.

\subsection{Relevance Verifier}
\label{sec:verifier}
By integrating heterogeneous data streams and internalizing cognitive reasoning, the Relevance Verifier delivers precise relevance assessments.

\subsubsection{Multi-Source Data Integration}

Unlike conventional relevance models constrained by sparse metadata, our architecture synthesizes heterogeneous signals to reconstruct a granular context. On the \textit{item side}, we incorporate OCR text from product detail pages, curated high-quality reviews, price, historical sales velocity, and store/brand reputation. OCR is performed offline and indexed alongside other item fields, incurring no per-query cost at serving time. On the \textit{user side}, we leverage features such as geolocation to ground the judgment in the querying context.

\subsubsection{Internalizing Reasoning via CoT}

We augment the model's cognitive capabilities via multi-instruction SFT enriched with CoT data, predicting a binary relevance label with a concise rationale to balance deductive capability with latency. The supervision corpus comprises $\sim\!800{,}000$ query-item pairs, stratified-sampled from production logs across the four complexity buckets in Section~\ref{sec:complexity} and across top-level Taobao categories in proportion to traffic. Annotation was performed by $\sim\!10$ trained annotators under a unified guideline; each pair was double-labeled, with disagreements ($\sim\!8\%$) adjudicated by a senior annotator, yielding Cohen's $\kappa = 0.81$ on a $2{,}000$-pair audit set. The final positive-to-negative ratio is $\sim\!1{:}1.4$, rebalanced via targeted negative mining. CoT rationales are generated by prompting DeepSeek-R1 \emph{without exposure to the human label}; we retain only those whose verdict agrees with the annotation. A $1{,}000$-sample spot-check confirmed that over $92\%$ of retained rationales are factually consistent with the label.

\subsubsection{Engineering for High-Throughput Deployment}

To reconcile the Verifier's computational intensity with strict online latency, we employ sequential batch processing, since per-item assessments are independent and users' scrolling naturally aligns with batch-wise filtering. This drastically reduces peak QPS, stabilizing at $\sim\!150$ during peak hours. We further design an \textit{adaptive pagination} strategy: if the Verifier filters out a significant portion of first-page items, the system automatically fetches and merges the next page, ensuring a full-page display without explicit user interaction.

On the serving stack, both the Expander and Verifier are deployed with INT8 weight quantization and tensor parallelism ($\mathrm{TP}{=}2$) per replica; combined with request-level continuous batching, this keeps end-to-end latency within the $\sim$290~ms P99 budget of Section~\ref{sec:method_overview} under production load. As detailed hardware footprint and per-query cost are commercially sensitive, we instead report behaviorally interpretable indicators---peak QPS and sustained-load latency---which are the operationally meaningful signals for practitioners.

\section{Experiments}

We conduct extensive experiments to evaluate the effectiveness of LEAPS, organized around the following research questions:

\begin{enumerate}[label=\textbf{RQ\arabic*:}, leftmargin=*, align=left, labelsep=0.6em]
    \item \textbf{(Overall effectiveness)} Does LEAPS consistently outperform strong baselines on both industrial and public retrieval backends?
    \item \textbf{(Architectural soundness)} Are the dual-plugin architecture and each Expander training stage individually necessary?
    \item \textbf{(Design motivation)} Are the gains of LEAPS concentrated on complex-constraint queries, the regime it is designed for?
    \item \textbf{(Online performance)} Does LEAPS deliver significant improvements in a real-world serving environment?
\end{enumerate}

\subsection{Experimental Setup}
\label{sec:setup}

\paragraph{Datasets.}
We evaluate LEAPS on two complementary datasets covering an industrial production setting and a reproducible academic benchmark. \textit{Taobao Internal} consists of $10{,}000$ conversational-style queries randomly sampled from Taobao AI Search logs; for each query, candidate items returned by every evaluated method are independently judged by trained annotators, yielding query--item relevance pairs for end-to-end metric computation. \textit{ESCI}~\cite{DBLP:journals/corr/abs-2206-06588} is the public Amazon Shopping Queries dataset; we use the English subset, focusing on the conversational-style query distribution that aligns with the multi-constraint setting LEAPS targets, and follow common e-commerce practice in treating \texttt{E} (Exact) and \texttt{S} (Substitute) labels as positive. We use ESCI as a reproducible black-box backend for portability evaluation rather than to compete with its leaderboard rerankers, which are pointwise query--item scorers and are not directly comparable as end-to-end pipelines coupling adaptive broadening with set-level verification.

\paragraph{Baselines.}
All methods are evaluated end-to-end on top of the same retrieval backend per dataset: the Taobao production retrieval system for \textit{Taobao Internal}, and a BM25\,+\,BGE hybrid index built on the ESCI catalog as a reproducible black-box engine. We compare against five baselines:
(i) \textbf{Production / Hybrid Retrieval}, the raw backend without any rewriting or filtering plugin;
(ii) \textbf{Query2Doc}~\cite{DBLP:conf/emnlp/WangYW23}, a representative LLM-based query rewriting baseline that prompts an LLM to generate pseudo-documents concatenated with the original query for retrieval, applied here on the same backend without a downstream filter;
(iii) \textbf{MiniELM}~\cite{DBLP:conf/acl/NguyenMYAC25}, a recent lightweight, adaptive query rewriting framework for e-commerce search that combines distillation, SFT and RL, evaluated under its standard configuration on each dataset following the authors' released setup;
(iv) \textbf{ICL LLM Pipeline}, which couples an ICL rewriter and an ICL filter on the same base model as LEAPS. Each is prompted with eight demonstrations spanning the four complexity buckets in Section~\ref{sec:complexity}, serving as a prompt-engineered upper bound without dedicated training;
(v) \textbf{Claude Opus 4}, which replaces the LEAPS Expander and Verifier with Anthropic's Claude Opus 4 prompted with the same eight-shot demonstration set as the ICL LLM Pipeline for query rewriting and relevance filtering.

\paragraph{Metrics.}
For offline evaluation we report Precision@20, Recall@20, and F1@20 against expert relevance judgments, where each method's final candidate pool is truncated to the top 20 items for a controlled, comparable assessment. We treat F1@20 as the primary metric, since it directly reflects the ``Broaden-and-Refine'' objective of jointly optimizing coverage and precision rather than trading one for the other. Crucially, the expert judgments used to compute P/R/F1@20 throughout Sections~\ref{sec:overall}--\ref{sec:complexity} are produced \emph{independently of the Verifier}; the Verifier appears only as a training-time component (Algorithm~\ref{alg:sft_posterior} line 9 and the RL reward), so the offline metrics are not circular with respect to its outputs. For online evaluation we report four complementary metrics defined in Section~\ref{sec:online}: Click-Through Rate (CTR), Exposures per Session (EPS), Clicks per Session (CPS), and Sufficient Result Rate (SRR).

\paragraph{Implementation Details.}
To facilitate reproducibility, we report all training and inference configurations in detail below. We adopt Qwen3-14B~\cite{qwen3} as the base model for both the Query Expander and the Relevance Verifier. The Expander is trained in three stages---inverse-data-augmentation SFT, posterior-knowledge SFT, and diversity-aware RL with the Effective Relevance reward---while the Verifier is trained via multi-instruction SFT with CoT rationales on approximately 800K manually labeled query--item pairs. For RL, we use GRPO within the ROLL framework~\cite{wang2025reinforcement}. For ESCI, we index the full English catalog ($\sim$1.2M items) with two parallel retrievers under standard configurations: BM25 (Anserini defaults) over concatenated \texttt{title}/\texttt{description}/\texttt{bullet\_point}, and \texttt{BAAI/bge-base-en-v1.5} with cosine similarity (FAISS HNSW). Each retriever returns top-200 candidates, fused via Reciprocal Rank Fusion into a top-100 pool truncated to top-20 for evaluation; the same configuration is applied uniformly across all methods on ESCI to ensure backend parity.

\subsection{Offline Evaluation (RQ1)}
\label{sec:overall}

Table~\ref{tab:main_results} reports Precision@20, Recall@20, and F1@20 against expert judgments on \textit{Taobao Internal} and \textit{ESCI}.

\begin{table}[htbp]
\centering
\caption{End-to-end comparison on \textit{Taobao Internal} (production retrieval) and \textit{ESCI} (BM25\,+\,BGE hybrid retrieval). All methods share the same backend per dataset.}
\label{tab:main_results}
\setlength{\tabcolsep}{4.0pt}
\renewcommand{\arraystretch}{1.15}
\begin{tabular}{llccc}
\toprule
\textbf{Dataset} & \textbf{Method} & \textbf{P@20} & \textbf{R@20} & \textbf{F1@20} \\
\midrule
\multirow{6}{*}{\shortstack[l]{\textit{Taobao}\\\textit{Internal}}}
 & Production Retrieval        & 0.62 & 0.58 & 0.60 \\
 & Query2Doc                   & 0.54 & 0.65 & 0.59 \\
 & MiniELM                     & 0.58 & 0.68 & 0.62 \\
 & ICL LLM Pipeline$^{\dagger}$ & 0.68 & 0.69 & 0.69 \\
 & Claude Opus 4$^{\S}$ & 0.71 & 0.73 & 0.72 \\
 & \textbf{LEAPS (Ours)}       & \textbf{0.79} & \textbf{0.82} & \textbf{0.81} \\
\midrule
\multirow{6}{*}{\shortstack[l]{\textit{ESCI}\\\textit{(Public)}}}
 & Hybrid Retrieval$^{\ddagger}$    & 0.55 & 0.51 & 0.53 \\
 & Query2Doc                        & 0.51 & 0.59 & 0.55 \\
 & MiniELM                          & 0.54 & 0.63 & 0.58 \\
 & ICL LLM Pipeline$^{\dagger}$ & 0.62 & 0.61 & 0.62 \\
 & Claude Opus 4$^{\S}$ & 0.66 & 0.66 & 0.66 \\
 & \textbf{LEAPS (Ours)}            & \textbf{0.72} & \textbf{0.74} & \textbf{0.73} \\
\bottomrule
\end{tabular}
\vspace{2pt}
\begin{flushleft}
\footnotesize
$^{\dagger}$ Eight-shot ICL rewriter + retrieval + eight-shot ICL filter, sharing the base model with LEAPS; the only difference is the absence of dedicated training.\quad
$^{\ddagger}$ BM25\,+\,BGE hybrid retrieval over the ESCI catalog.\quad
$^{\S}$ Claude Opus 4 is used with the same eight-shot prompting protocol as the ICL LLM Pipeline for query rewriting and Verifier filtering.
\end{flushleft}
\end{table}

Across both datasets, LEAPS achieves the highest F1@20 by a clear margin, and the improvement decomposes consistently into gains on \emph{both} precision and recall rather than trading one for the other. This is a direct empirical reflection of the Broaden-and-Refine design: the Expander enlarges the candidate set in a system-friendly manner, and the Verifier prunes noise through grounded reasoning. Gains concentrated in F1@20 therefore carry stronger evidential weight than gains in any individual component.

The margin over MiniELM stands out on both backends. We attribute this to two design choices: MiniELM optimizes rewrites individually, whereas LEAPS captures set-level utility via its diversity-aware reward; and MiniELM lacks a downstream semantic gate, leaving the precision cost of broadening to the user. Query2Doc, expanding queries via LLM-generated pseudo-documents without system-aware training or downstream filtering, lifts recall but degrades precision---generic LLM rewriting does not capture e-commerce-specific constraint dynamics. The ICL LLM Pipeline isolates the contribution of our training paradigm versus raw LLM capability and prompting. Since it shares the base model and uses few-shot demonstrations matched to our query complexity buckets, the residual gap is attributable to the three-stage Expander training and the multi-instruction CoT Verifier training, indicating that in-context prompting alone is insufficient for production-grade conversational search. Claude Opus 4 reinforces this from another angle: with a strictly stronger backbone and the same eight-shot demonstrations, it still trails LEAPS---gains stem from our training paradigm, not base-model capacity.

The consistency across two structurally distinct backends---a decade-optimized industrial system and a minimal hybrid index built solely for this evaluation---directly evidences the non-intrusive, portable nature of LEAPS, in contrast to prior work whose benefits are entangled with backend-specific signals such as click logs or proprietary embeddings.

\subsection{Ablation Study (RQ2)}
\label{sec:ablation}

We next ask whether each design choice within the framework is individually justified. Table~\ref{tab:ablation} reports a structured ablation organized into three groups, each targeting a distinct question.

\begin{table}[htbp]
\centering
\caption{Component-wise ablation of LEAPS, organized into three groups: architectural composition, Expander training stages, and Verifier design choices.}
\label{tab:ablation}
\setlength{\tabcolsep}{6pt}
\renewcommand{\arraystretch}{1.15}
\begin{tabular}{lccc}
\toprule
\textbf{Variant} & \textbf{P@20} & \textbf{R@20} & \textbf{F1@20} \\
\midrule
\multicolumn{4}{l}{\textit{Architectural Integrity}} \\
\quad Expander only                & 0.48 & 0.85 & 0.61 \\
\quad Verifier only                & 0.84 & 0.42 & 0.56 \\
\midrule
\multicolumn{4}{l}{\textit{Expander Training}} \\
\quad w/o Stage 1 (Data Aug.)      & 0.74 & 0.77 & 0.75 \\
\quad w/o Stage 2 (Posterior SFT)  & 0.66 & 0.69 & 0.67 \\
\quad w/o Stage 3 (RL)             & 0.73 & 0.72 & 0.72 \\
\midrule
\multicolumn{4}{l}{\textit{Verifier Design}} \\
\quad w/o CoT reasoning            & 0.71 & 0.74 & 0.72 \\
\quad w/o multi-source features$^{\P}$ & 0.69 & 0.73 & 0.71 \\
\midrule
\textbf{LEAPS (Full)}              & \textbf{0.79} & \textbf{0.82} & \textbf{0.81} \\
\bottomrule
\end{tabular}
\vspace{2pt}
\begin{flushleft}
\footnotesize
$^{\P}$ Only product title and attributes are fed to the Verifier; OCR, reviews, sales, and store/brand signals are removed.
\end{flushleft}
\end{table}

\paragraph{Architectural integrity.}
The Expander and Verifier address orthogonal failure modes---precision and recall---and only their composition reaches a usable operating point (see Table~\ref{tab:ablation}).

\paragraph{Expander training stages.}
Removing Stage 1 (Inverse Data Augmentation) degrades F1@20, confirming that warm-up alignment between colloquial language and structured product indices is a prerequisite. Removing Stage 2 (Posterior-Knowledge SFT) produces the largest single-stage drop, validating our central claim that posterior search yield---not a priori semantic similarity---is the right supervision signal for industrial query rewriting. Removing Stage 3 (Diversity-aware RL) preserves per-rewrite quality but collapses the rewrite set into homogeneous variants, stalling incremental coverage. The three stages thus contribute along distinct axes: lexical alignment, system-aware utility, and set-level diversity.

\paragraph{Verifier design.}
Removing CoT (training the Verifier to predict binary labels without rationale) drops F1@20 mainly on multi-constraint queries, where surface pattern matching is insufficient. Removing multi-source features---keeping only title and structured attributes, dropping OCR, reviews, sales, and store/brand---causes comparable degradation, concentrated on queries whose constraints are absent from the title (e.g., materials shown only in detail-page images). An internal analysis (omitted for space) shows OCR and curated reviews drive the largest individual gains (OCR on material/visual constraints, reviews on usage-scenario constraints), while sales and store/brand mainly improve long-tail precision.

\subsection{Robustness to Query Complexity (RQ3)}
\label{sec:complexity}

LEAPS is motivated by a specific claim: traditional retrieval degrades sharply as queries accumulate constraints, and a Broaden-and-Refine plugin can recover performance precisely in this regime. To test this, we stratify evaluation queries by constraint count, prompting Qwen3-32B to enumerate constraint clauses (category, material, price range, scenario, negation) and bucketing queries into \textit{Simple} (1), \textit{Moderate} (2), \textit{Complex} (3), and \textit{Heavy} ($\geq$4). Manual re-annotation on a random sample of $200$ queries shows $\sim$92\% bucket-level agreement with the Qwen3-32B output, confirming the stratification is sufficiently stable. The distribution of production conversational-style traffic across the four buckets is broadly consistent with that of the offline evaluation set, so the bucket-wise gains reported below are representative at the population level rather than concentrated in a low-traffic tail.

\begin{table}[htbp]
\centering
\caption{F1@20 across query complexity buckets, where queries are stratified by attribute constraint count: Simple (1), Moderate (2), Complex (3), and Heavy ($\geq$4).}
\label{tab:complexity}
\setlength{\tabcolsep}{3.6pt}
\renewcommand{\arraystretch}{1.15}
\begin{tabular}{lcccc}
\toprule
\textbf{Method} & \textbf{Simple} & \textbf{Moderate} & \textbf{Complex} & \textbf{Heavy} \\
\midrule
Raw Retrieval                  & 0.79 & 0.64 & 0.47 & 0.32 \\
Query2Doc                      & 0.80 & 0.66 & 0.49 & 0.36 \\
MiniELM                        & 0.81 & 0.67 & 0.52 & 0.38 \\
ICL LLM Pipeline               & 0.83 & 0.72 & 0.60 & 0.46 \\
Claude Opus 4                  & 0.85 & 0.75 & 0.64 & 0.52 \\
\textbf{LEAPS (Ours)}          & \textbf{0.86} & \textbf{0.82} & \textbf{0.77} & \textbf{0.70} \\
\bottomrule
\end{tabular}
\end{table}

Table~\ref{tab:complexity} shows that the relative gain of LEAPS over the retrieval baseline grows monotonically with constraint count. On single-constraint queries, where the baseline already operates near its design envelope, LEAPS offers only a modest improvement. As constraints accumulate, the baseline deteriorates sharply---reflecting the over-constraint failure mode in which rigid conjunctive matching either returns zero results or surfaces only partial matches---while LEAPS degrades far more gracefully. Query2Doc, MiniELM, the ICL LLM Pipeline, and Claude Opus 4 all track the baseline with comparatively flat relative gains, indicating that neither prior rewriting methods nor a substantially stronger general-purpose LLM under matched eight-shot prompting specifically addresses the complexity dimension. This matches the design: the Expander's posterior-aware relaxation matters most when constraints are numerous enough to trigger over-constraint failures, and the Verifier's holistic reasoning matters most when the broadened pool contains many partial matches.

\subsection{Online Performance (RQ4)}
\label{sec:online}

We complement the offline analysis with a one-month online A/B test on a representative slice of production traffic on Taobao AI Search in July 2025, well outside any major promotional period. The control group serves traffic through the default pipeline without LEAPS, while the treatment group serves the same traffic through LEAPS. We track four metrics jointly characterizing the conversational shopping experience: \textit{Click-Through Rate} (CTR, impression-level clicks), \textit{Exposures per Session} (EPS, breadth), \textit{Clicks per Session} (CPS, actionable conversion), and \textit{Sufficient Result Rate} (SRR, fraction of queries returning $\geq$10 relevant items after the full pipeline). SRR speaks directly to the zero-result dilemma motivating this work. All four metrics are statistically significant at $p$ < 0.001 over the observation window.

\begin{table}[htbp]
\centering
\caption{Online comparison on Taobao AI Search. \textit{Control} denotes the default pipeline without LEAPS.}
\label{tab:online}
\setlength{\tabcolsep}{6pt}
\renewcommand{\arraystretch}{1.15}
\begin{tabular}{lccc}
\toprule
\textbf{Metric} & \textbf{Control} & \textbf{LEAPS} & \textbf{Rel.\ Gain} \\
\midrule
CTR              & 9.4\%   & 11.9\%  & +27.1\%  \\
EPS              & 11.3    & 12.8    & +13.3\%  \\
CPS              & 0.82    & 1.18    & +43.9\%  \\
SRR              & 77.0\%  & 91.4\%  & +18.7\%  \\
\bottomrule
\end{tabular}
\end{table}

Table~\ref{tab:online} reports the comparison. The A/B runs on conversational-style queries, the regime where Section~\ref{sec:complexity} documented a sharp degradation of the retrieval baseline; the lifts are therefore the deployment-time manifestation of that offline gap, not a uniform shift over Taobao traffic. The CTR gain indicates that surfaced items are better aligned with user intent---a direct consequence of the Verifier's semantic gatekeeping. EPS grows modestly because an intentional per-page surface-area cap bounds rendering cost, masking the Expander's raw broadening capacity. CPS shows the largest lift, reflecting that breadth and per-item quality compound multiplicatively into session-level engagement gains beyond what either factor delivers in isolation. The SRR climb confirms that the Expander's posterior-aware constraint relaxation resolves the zero-result dilemma in production. These results are consistent with the complexity-stratified analysis: gains concentrate on multi-constraint queries where the baseline operates far from saturation.

\begin{figure}[htbp]
  \centering
  \includegraphics[width=0.5\linewidth]{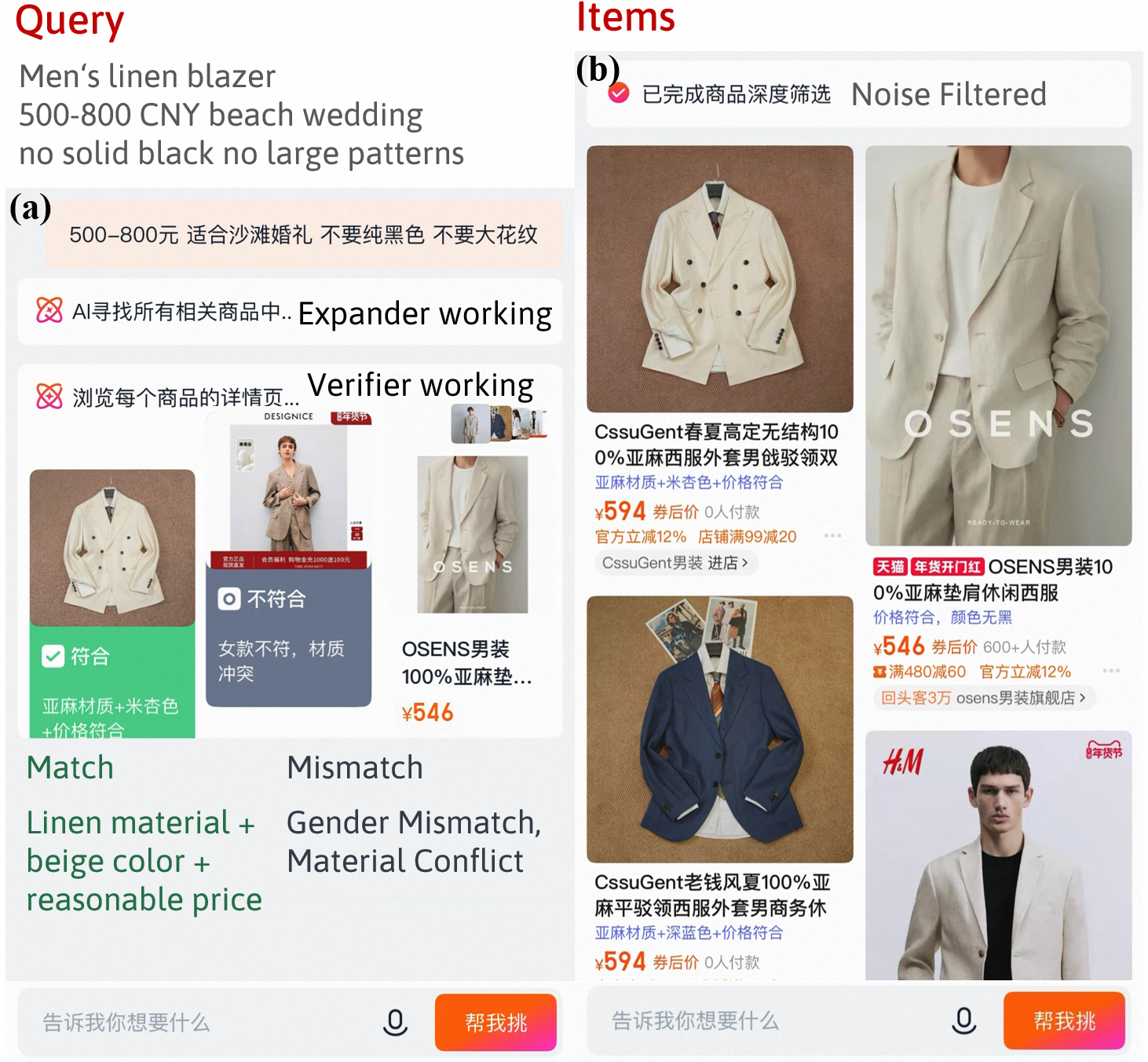}
  \caption{A production case from the Taobao AI Search Chinese mobile interface on a five-constraint query: (a) Verifier judgments with concise rationales overlaid on candidate items; (b) the final dual-column feed after noise filtering.}
  \label{fig:case-study}
\end{figure}

LEAPS has been fully deployed on Taobao AI Search since August 2025, serving hundreds of millions of monthly users. Figure~\ref{fig:case-study} shows a production trace for the query \textit{``men's linen blazer, 500--800 CNY, beach wedding, no solid black, no large patterns''}---a query with five simultaneous constraints. The Expander broadens scope into a candidate pool; the Verifier then inspects each item against the full constraint set and emits a binary judgment with concise rationale, yielding a noise-filtered feed. An audit of $200$ randomly sampled multi-constraint dialogues finds LEAPS produces a fully compliant first page in $84\%$ of cases versus $41\%$ for control, confirming that the gains in Table~\ref{tab:online} are population-level rather than cherry-picked.

\section{Conclusion}
\label{sec:conclusion}

We presented LEAPS, a non-intrusive plugin bridging conversational queries and traditional search via a Broaden-and-Refine paradigm: a Query Expander---trained with inverse data augmentation, posterior-knowledge SFT, and diversity-aware RL---resolves zero-result queries, while a Relevance Verifier with multi-source context and internalized CoT mitigates decision overload. Offline and one-month online A/B experiments validate LEAPS, deployed on Taobao AI Search since August 2025 to hundreds of millions of monthly users. Two limitations point to future work: equal weighting of rewrites within the budget motivates confidence-aware allocation; and residual low-result cases on idiosyncratic queries with rare entities suggest coupling Expander broadening with retrieval-side index expansion.

\bibliographystyle{unsrtnat}
\bibliography{references}  %%% Uncomment this line and comment out the ``thebibliography'' section below to use the external .bib file (using bibtex) .




% 	\bibitem{keshet2016prediction}
% 	Keshet, Renato, Alina Maor, and George Kour.
% 	\newblock Prediction-Based, Prioritized Market-Share Insight Extraction.
% 	\newblock In {\em Advanced Data Mining and Applications (ADMA), 2016 12th International 
%                       Conference of}, pages 81--94,2016.

% \end{thebibliography}

\end{document}